\documentclass[11pt]{article}

\usepackage{graphicx}
\usepackage{amsmath,amsfonts,amssymb}
\usepackage{fullpage}
\usepackage[colorlinks,citecolor=blue]{hyperref}

\newcommand\eqnref[1]{(\ref{#1})}
\newcommand\figref[1]{Fig.~\ref{#1}}

\newcommand\sectref[1]{Section~\ref{#1}}

\newcommand{\iu}   {\mathrm{i}}     

\newcommand{\Omegarm}   {\mathrm{\Omega}}

\newcommand{\bfH}   {\mathbf{H}}
\newcommand{\bfB}   {\mathbf{B}}
\newcommand{\bfE}   {\mathbf{E}}
\newcommand{\bfD}   {\mathbf{D}}
\newcommand{\bfb}   {\mathbf{b}}
\newcommand{\bfd}   {\mathbf{d}}
\newcommand{\bfe}   {\mathbf{e}}
\newcommand{\bfh}   {\mathbf{h}}
\newcommand{\bfp}   {\mathbf{p}}

\newcommand{\bfk}   {\mathbf{k}}
\newcommand{\bfm}   {\mathbf{m}}

\newcommand{\bfr}   {\mathbf{r}}
\newcommand{\bfv}   {\mathbf{v}}
\newcommand{\bfj}   {\mathbf{j}}
\newcommand{\bfq}   {\mathbf{q}}
\newcommand{\bfw}   {\mathbf{w}}
\newcommand{\bfK}   {\mathbf{K}}
\newcommand{\bfQ}   {\mathbf{Q}}
\newcommand{\bfF}   {\mathbf{F}}
\newcommand{\epseff}  {\epsilon_{\mathrm{eff}}}
\newcommand{\mueff}   {\mu_{\mathrm{eff}}}

\newcommand{\figurewidth}   {0.85\linewidth}

\begin{document}

\title{From Whitney Forms to Metamaterials: a Rigorous Homogenization Theory}

\author{Igor Tsukerman\\
Department of Electrical and Computer Engineering,\\
The University of Akron, OH 44325-3904, USA\\
igor@uakron.edu
}


\maketitle

\textbf{Abstract}.  A rigorous homogenization theory of metamaterials -- artificial
periodic structures judiciously designed to control the propagation of electromagnetic waves --
is developed. All coarse-grained fields are unambiguously defined and effective
parameters are then derived without any heuristic assumptions.
The theory is an amalgamation of two concepts: Smith \& Pendry's physical insight
into field averaging and the mathematical framework of Whitney-Nedelec-Bossavit-Kotiuga interpolation.
All coarse-grained fields are defined via Whitney forms
and satisfy Maxwell's equations exactly. The new approach is illustrated with several
analytical and numerical examples and agrees well with the established results
(e.g. the Maxwell-Garnett formula and the zero cell-size limit) within the range
of applicability of the latter. The sources of approximation
error and the respective suitable error indicators are clearly identified,
along with systematic routes for improving the accuracy further.
The proposed approach should be applicable in areas beyond metamaterials
and electromagnetic waves -- e.g. in acoustics and elasticity.

\section{Introduction}\label{sec:Intro}
%
Electromagnetic and optical metamaterials are periodic structures with
features smaller than the vacuum wavelength, judiciously designed
to control the propagation of waves. Typically, resonance elements
(variations of split-ring resonators) are included to produce nontrivial
and intriguing effects such as backward waves and negative refraction, cloaking, slow light
(``electromagnetically induced transparency''), and more; see
\cite{Alu07,Buell06,Ikonen06,Papasimakis08,Schurig06,Smith00,Yousefi07,Zhang08,Ziolkowski06},
to name just a few representative publications, and references there.

To gain insight into the behavior of such artificial structures and to be able to design useful devices,
one needs to approximate a given metamaterial by an effective medium
with dielectric permittivity $\epseff$ and magnetic permeability $\mueff$ or, in more general
cases where magnetoelectric coupling may exist, with a $6 \times 6$ parameter matrix.
A variety of approaches have been explored in the literature (e.g.
\cite{Sarychev-Shalaev-book07,Silveirinha07,Simovski09,Simovski-Tretyakov10,Smith06,Tretyakov-book03},
again to name just a few publications),
with notable accomplishments in designing novel and interesting devices and structures,
as cited above. However,
%
%
%
the existing approaches are mostly heuristic, and there is a clear need for
a consistent and rigorous theory -- rigorous in the sense that all ``macroscopic''
(coarse-grained) fields are unambiguously and precisely defined,
giving rise to equally well defined effective parameters.
%

It is the main objective of this paper to put forward such a theory.
The methodology advocated here is an amalgamation of two very different lines of thinking:
one relatively new and driven primarily by physical insight, and the other one
well established and mathematically rigorous. The physical insight
is due to Pendry \& Smith \cite{Smith06}, who prescribed different averaging procedures
for the $\bfh$ and $\bfb$ fields (and similarly for the $\bfe$ and $\bfd$
fields). The practical results have been excellent, and yet it has remained a bit of a mystery
why, say, $\bfh$ and $\bfb$ must be averaged differently. The justification
for that in \cite{Smith06} and other publications comes from the analogy
with staggered grid approximations in finite difference methods,
but it is unclear why physics should be subordinate to numerical methods
and not the other way around.

\emph{Remark}. Throughout the paper, small letters $\bfb, \bfe, \bfd$, etc.,
will denote the ``microscopic'' -- i.e. true physical -- fields that in general
vary rapidly as a function of coordinates. Capital letters will refer to
smoother fields, to be defined precisely later, that vary on the scale coarser than the lattice
cell size.

The second root of the proposed methodology is the mathematical framework
developed by Whitney in the middle of the 20th century \cite{Whitney57}. A few
decades later Whitney's theory was discovered to be highly relevant in computational
electromagnetism, due primarily to the work of Nedelec, Bossavit and Kotiuga
\cite{Bossavit88a,Bossavit98,Kotiuga85,Nedelec80,Nedelec86}. It should be emphasized,
however, that in the present paper the Whitney-Nedelec-Bossavit-Kotiuga (WNBK)
framework is used \emph{not} for computational
purposes but to define, analytically, the coarse-grained fields.

The end result of combining WNBK interpolation with Smith \& Pendry's insight
is a mathematically and physically consistent model
that is rigorous, general (e.g., applicable to magnetoelectric coupling)
and yet simple enough to be practical. The theory is supported
by a number of analytical and numerical case studies and is consistent
with the existing theories and results (e.g. with the Maxwell-Garnett mixing formula
and with the zero cell-size limit) within the ranges of applicability of the latter.
Approximations that have to be made, and the respective sources of error,
are clearly identified. Several routes for further accuracy improvement
are apparent from the theoretical analysis.
%
\section{Some Pitfalls}\label{sec:Pitfalls}
%
Prior to discussing what needs to be done, it is instructive to review
what \emph{not} to do. Some averaging procedures appear to be quite natural and yet
upon closer inspection turn out to be flawed.
Physically valid alternatives are introduced in the subsequent sections.

\textbf{Passing to the limit of the zero cell size.}
A distinguishing feature of the homogenization problem for metamaterials
is that the cell size $a$, while smaller than
the vacuum wavelength $\lambda_0$ at a given frequency $\omega$, is not vanishingly
small. A typical range in practice is $a \sim 0.1 - 0.3 \lambda_0$. Therefore classical
homogenization procedures valid for $a \rightarrow 0$ -- e.g. Fourier
\cite{Sjoberg05-149} or two-scale analysis \cite{Bensoussan78} -- have limited or no applicability here.
Independent physical \cite{Merlin09} and mathematical \cite{Tsukerman08}
arguments show that the finite cell size is a \emph{principal} limitation
rather than just a constraint of fabrication. If the cell size is reduced
relative to the vacuum wavelength, the nontrivial physical effects
(e.g. ``artificial magnetism'') ultimately disappear,
provided that the intrinsic dielectric permittivity $\epsilon$ of the materials
remains unchanged. On physical grounds \cite{Tsukerman08}, this can be explained
by the operating point on the normalized Bloch band diagram falling
on the ``uninteresting'' acoustic branch.

\textbf{Mollifiers.} The classical approach to defining the macroscopic fields
is via convolution with a smooth mollifier function (e.g. Gaussian-like)
\cite{Russakoff70}. This necessitates an intermediate scale for the mollifier,
much coarser than the cell size but still much finer than the wavelength in the material.
For natural materials, this requirement is easily fulfilled because
the cell size is on the order of molecular dimensions and much smaller
than the wavelength. In contrast, for metamaterials the cell size
is typically $\sim$0.1 -- 0.3 of the wavelength, and no intermediate scale
is available for the mollifier.

\begin{figure}
  \centering
  \includegraphics[width=0.75\linewidth]{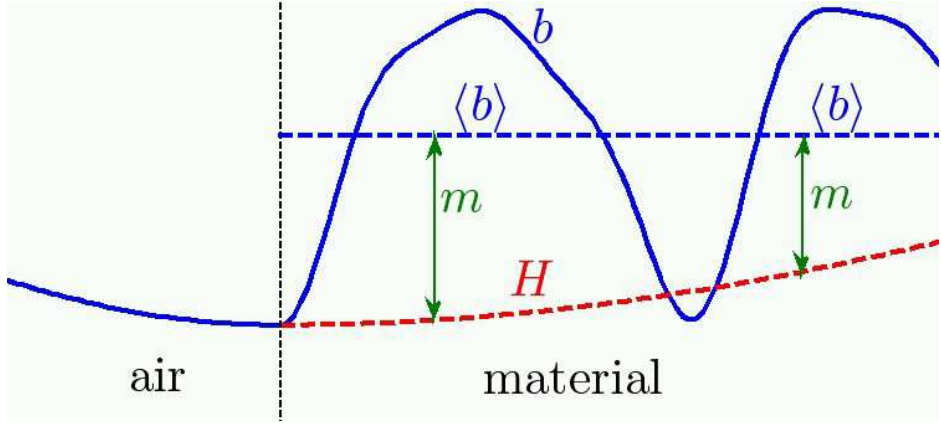}\\
  \caption{The field and magnetization in the direction normal to the interface.}\label{fig:b-vs-n}
\end{figure}

\textbf{Averaging over the cell.} The following argument shows that
simple cell-averaging of the fields is for metamaterials inadequate.
To fix ideas, consider a simplified picture first. A qualitative behavior
of a tangential component of the microscopic $\bfb$-field in the direction normal
to the surface is shown in \figref{fig:b-vs-n}. For our purposes at this point,
the precise field distribution is unimportant; one may have in mind, for example,
a single Bloch wave moving away from the surface, even though later on it will be important
to consider a \emph{superposition} of Bloch waves.

Clearly, the average field $\bfB_0 = \langle \bfb \rangle$ over the cell
is not in general equal to the field
$\bfb(0+)$ inside the material immediately at the boundary. Yet it is $\bfb(0+)$
that couples with the field in the air: $\bfb(0+)= \bfb(0-)$, where $\bfb(0-)$ is the field
in the air immediately at the boundary. (Intrinsically nonmagnetic materials
are assumed throughout.) The classical boundary condition is recovered if an auxiliary
$\bfH$ field is introduced in such a way that $\bfH(0+) = \bfb(0+)$; magnetization then is
$4\pi \bfm = \bfb - \bfH$, as schematically shown in the figure. In other words, it is the difference
between the cell-averaged field $\bfb$ and its value at the boundary
that ultimately defines magnetization and the permeability.

This picture, albeit simplified (in particular, it ignores
a complicated surface wave at the interface \cite{Simovski09,Markel-private10}),
does serve as a useful starting point for a proper physical definition
of field averages and effective material parameters.

Since the cell-averaged field is, for a non-vanishing cell size, not generally equal
to its boundary value and does not satisfy the proper continuity condition
at the interface, the use of such fields in a model would result in
nonphysical equivalent electric / magnetic currents on the surface (jumps of a tangential
component of the magnetic / electric field), nonphysical electric/magnetic surface charges
(jumps of the normal component of $\bfD$ or $\bfB$), and incorrect reflection/transmission
conditions at the boundary.
It is true that, as a ``zero-order'' approximation, the cell-averaged field is approximately
equal to its pointwise value; however, equating them means ignoring the variation
of the fields over the cell and hence throwing away the very physical effects
under investigation.

\textbf{``Magnetic dipole moment per unit volume.''} This textbook definition
of magnetization
turns out to be flawed as well.
Simovski \& Tretyakov \cite{Simovski-Tretyakov10} give a counterexample
for a system of two small particles, but in fact their argument is general.
Suppose that a large volume of a metamaterial has been in some way homogenized
and is now represented, to an acceptable level of approximation, by effective parameters
$\mueff$, $\epseff$. Consider then a standing electromagnetic wave in this material
(as produced e.g. in a cavity or by reflection off a mirror).
It is well known that the electric and magnetic fields in a lossless standing wave are shifted
by one quarter of the wavelength. At a node of the electric field, i.e. at a point in space
where the electric field is zero, the magnetic flux density and hence the magnetization
are \emph{maximum}. Yet the zero electric field at the node implies zero currents (both polarization
and conduction) and hence zero magnetic dipole moment (as there is no spin-related intrinsic moment
by assumption). This inconsistency shows that magnetization cannot be defined
as the dipole moment per unit volume.


\textbf{Bulk parameters.} It is known that even for a homogeneous isotropic \emph{infinite} medium
the pair of parameters $\epsilon$ and $\mu$ are not defined uniquely.
Indeed, the total microscopic current $\bfj$ can be split up -- in principle, fairly arbitrarily --
into the ``electric'' and ``magnetic'' parts, $\bfj = \partial_t \bfp + c \nabla \times \bfm$
\cite{Vinogradov02,Markel08,Markel-private10}.
The $\bfh$ field is then defined accordingly, as $\bfb - 4\pi \bfm$, giving rise to the respective value
of $\mueff$ that depends on the choice of $\bfm$. A more general ``Serdyukov-Fedorov''
transformation leaves Maxwell's equations invariant but changes the values of the material
parameters \cite{Bokut72,Bokut74,Vinogradov02}:

$$
  \bfd' = \bfd + \nabla \times \bfQ, ~~~
  \bfh' = \bfh + c^{-1} \partial_t \bfQ
$$
$$
  \bfb' = \bfb + \nabla \times \bfF, ~~~
  \bfe' = \bfe + c^{-1} \partial_t \bfF
$$
where $\bfQ$ and $\bfF$ are arbitrary fields (with a valid curl).
It is possible \cite{Agranovich84,Vinogradov02} to set $\mu = 1$,
in which case the dielectric function carries all relevant information
but becomes spatially dispersive (its transform depends
on the spatial Fourier vector $\bfk$).

Thus even for a homogeneous infinite medium it is only the product $\epsilon \mu$ that
is unambiguously defined, with its direct physical relation to phase velocity
$v_p = \sqrt{\epsilon \mu}$. The situation changes thoroughly when a material interface
(for simplicity, with air) is considered. Classical boundary conditions
for the tangential continuity of the $\bfH$ and $\bfE$ fields\footnote{These conditions
are local and much simpler than the ones arising in the model with $\mu \equiv 1$
\cite{Agranovich84,Vinogradov02}.} fix the ratio of the material parameters
via the intrinsic impedance $\eta = \sqrt{ \mu / \epsilon }$,
which, taken together with their product, identifies these two parameters separately
and uniquely.

It is clear, then, that any complete and rigorous definition
of the effective electromagnetic parameters of metamaterials must account for boundary effects.
%
%
%
\section{Coarse-Grained Fields}\label{sec:Coarse-grained-fields}
%
\subsection{The Guiding Principles}\label{sec:Guiding-principles}
%
Consider a periodic structure composed of materials that are assumed to be
(i) intrinsically nonmagnetic (which is true at sufficiently high frequencies
\cite{Landau84,Merlin09});
(ii) satisfy a linear local constitutive relation $\bfd = \epsilon \bfe$.
For simplicity, we assume a cubic lattice with cells of size $a$.

Maxwell's equations for the microscopic fields are, in the frequency domain and with the
$\exp(-\iu \omega t)$ phasor convention,
$$
 \nabla \times \bfe ~=~ \iu \omega c^{-1}\bfb
$$
$$
 \nabla \times \bfb ~=~ -\iu \omega c^{-1}\bfd
$$
The coarse-grained fields $\bfB$, $\bfH$, $\bfE$, $\bfD$ must
be defined in such a way that the boundary conditions are honored.
(As already noted, simple cell-averaging does not satisfy this condition.)
From the mathematical perspective, these fields must lie
in their respective functional spaces
\begin{equation}\label{eqn:EHBD-in-H-curl-H-div}
    \bfE, \bfH \in H(\mathrm{curl, \Omegarm}); ~~~
    \bfB, \bfD \in H(\mathrm{div, \Omegarm})
\end{equation}
where $\Omegarm$ is a domain of interest that for mathematical simplicity
is assumed finite. Symbol $\Omegarm$ will henceforth be dropped to shorten the notation.
Rigorous definitions of these functional spaces are available in the
mathematical literature (e.g. \cite{Monk03}). From the physical perspective,
constraints \eqnref{eqn:EHBD-in-H-curl-H-div} mean that
the $\bfE$, $\bfH$ fields possess a valid curl as a regular function
(not as a Schwartz distribution), while $\bfB$ and $\bfD$ have a valid divergence.
This implies, most importantly, tangential continuity of $\bfE$, $\bfH$
and normal continuity of $\bfB$ and $\bfD$ across material interfaces.
The fields in $H(\mathrm{curl})$ are also said to be \emph{curl-conforming},
and those in $H(\mathrm{div})$ to be \emph{div-conforming}.

In principle, any choice of a curl-conforming $\bfH$ field produces the respective
``magnetization'' $4\pi \bfm \equiv \bfb - \bfH$ and leaves Maxwell's equations intact.
However, most of such choices will result in technically valid but
completely impractical and arbitrary constitutive laws, with the ``material parameters''
depending more on the choice of $\bfH$ than on the material itself.

%

Below, I argue that construction of the coarse-grained fields via
the Whitney-Nedelec-Bossavit-Kotiuga (WNBK) interpolation has particular
mathematical and physical elegance, which leads to practical advantages
in the computation of fields in periodic structures.
%
\subsection{Background: WNBK Interpolation}\label{sec:WNBK-complex}
%
For a rigorous definition of the coarse-grained $\bfH$-field, we shall need an interpolatory
structure referred to in the literature as the ``Whitney complex'' \cite{Bossavit88a,Bossavit98};
however, acknowledgment of the seminal contributions of Nedelec, Bossavit and Kotiuga
\cite{Nedelec80,Bossavit88a,Kotiuga85} is quite appropriate and long overdue.
Still, the subscripts of quantities related to the Whitney complex will for brevity
be just `W' rather than `WNBK'.

WNBK complexes form a basis of modern finite element analysis with edge and facet elements.
However, our objective here is \emph{not} to develop a numerical procedure;
rather, the mathematical structure that has served so well in numerical analysis
is borrowed and applied to fields in metamaterial cells.

The original Whitney forms \cite{Whitney57} are rooted very deeply in differential geometry
and algebraic topology, and the interested reader can find a complete mathematical exposition
in the literature cited above. For the purposes of the present paper, we need a small subset of this
theory where the usual framework of vector calculus is sufficient. It will also
suffice to consider a reference cell as a cube $[-1, +1]^3$.
This shape can be transformed into an arbitrary rectangular parallelepiped by simple
scaling and, if necessary, to any hexahedron by a linear transformation as in
\cite{vanWelij85}.

We shall need two interpolation procedures: (i) given the circulations $[q]_\alpha \equiv
\int_\alpha \bfq \cdot \mathbf{dl}$ of a field $\bfq$ over the 12 edges of the cell
($\alpha = 1,2, \ldots, 12$), extend that field into the volume of the cell; and (ii)
given the fluxes $[[q]]_\beta \equiv \int_\beta \bfq \cdot \mathbf{dS}$ of a field $\bfq$
over the six faces of the cell ($\beta = 1,2, \ldots, 6$),
extend that field into the volume of the cell.
Single brackets denote line integrals over an edge; double brackets -- surface integrals
over the faces. Typically, item (i) will apply to the $\bfe$ and $\bfh$ fields and (ii) --
to the $\bfd$ and $\bfb$ fields.

\begin{figure}
  \centering
  \includegraphics[width=\figurewidth]{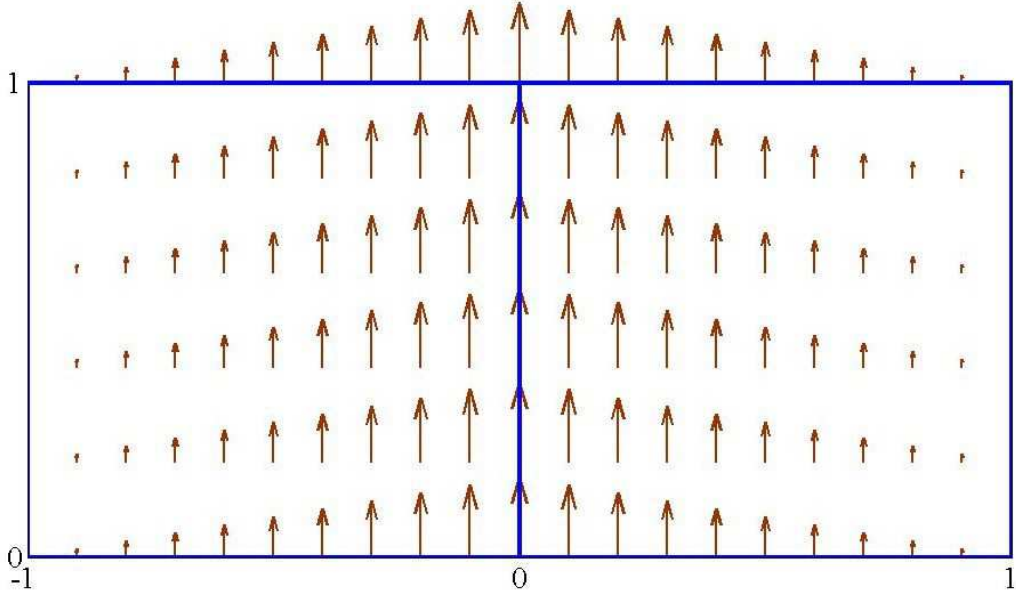}\\
  \caption{A 2D analog of the vectorial interpolation function $\bfw_\alpha$
  (in this case, associated with the central vertical edge shared by two adjacent cells).
  Tangential continuity of this function is evident from the arrow plot;
  its circulation is equal to one over the central edge and to zero
  over all other edges.}\label{fig:vector-edge-function}
\end{figure}

Consider an edge $\alpha$ along a $\xi$-direction, where $\xi$ is one of the coordinates $x, y, z$,
and let $\eta$ and $\tau$ be the other two coordinates, with $(\xi, \eta, \tau)$ being
a cyclic permutation of $(x, y, z)$ and hence a right-handed system.
The edge $\alpha$ is then formally defined as
$$
    -1 \leq \xi \leq 1; ~~~ \eta = \eta_\alpha; ~~ \tau = \tau_\alpha;
    ~~ \eta_\alpha, \tau_\alpha = \pm 1
$$
Associated with this edge is a vectorial interpolating function
\begin{equation}\label{eqn:hexahedral-edge-interpolant}
    \bfw_\alpha ~=~ \frac{1}{8} \, (1 + \eta_\alpha \eta) (1 + \tau_\alpha \tau)
    \, \hat{\xi}
\end{equation}
where the hat symbol denotes the unit vector in a given direction.
For illustration, a 2D analog of this vector function is shown in
\figref{fig:vector-edge-function}.

In 3D, there are 12 of such interpolating functions -- one per edge -- in the cell.
It is straightforward to verify that the edge circulations of these functions
have the Kronecker-delta property
\begin{equation}\label{eqn:hexahedral-edge-interpolant-Kronecker-delta}
    [\bfw_\alpha]_{\alpha'} ~=~ \delta_{\alpha \alpha'}
\end{equation}
Indeed, on edge $\alpha$ itself, $\eta_\alpha = \tau_\alpha = \eta = \tau = 1$;
hence the value of $\bfw_\alpha$ on its ``own'' edge is $\frac12$, which produces
unit circulation upon integration over the edge of length two ($[-1, +1]$).
For edges not along the $\xi$ direction, the circulation is obviously zero
because the vector function  $\bfw_\alpha$ is orthogonal to them.
For other edges in the $\xi$ direction, one of the coordinates $\eta$ or $\tau$
is equal to $-1$ and the respective factor in \eqnref{eqn:hexahedral-edge-interpolant-Kronecker-delta}
evaluates to zero, again producing a zero circulation.

The Kronecker-delta property guarantees that the interpolating functions are linearly
independent over the cell and span a 12-dimensional space of vectors that can all
be represented by interpolation from the edges into the volume of the cell:
\begin{equation}\label{eqn:hexahedral-edge-interpolation}
    \bfq ~=~ \sum_{\alpha=1}^{12} \, [\bfq]_\alpha \bfw_\alpha
\end{equation}
We shall call this 12-dimensional space $W_\mathrm{curl}$: the `W' honors Whitney
and `curl' indicates fields whose curl is a regular function
rather than a general distribution. This implies, in physical terms,
the absence of equivalent surface currents
and the tangential continuity of the fields involved. Any adjacent lattice cells
sharing a common edge will also share, by construction of interpolation
\eqnref{eqn:hexahedral-edge-interpolation},
the field circulation over that edge, which ensures the continuity of the tangential
component of the field across all edges.

\begin{figure}
  \centering
  \includegraphics[width=\figurewidth]{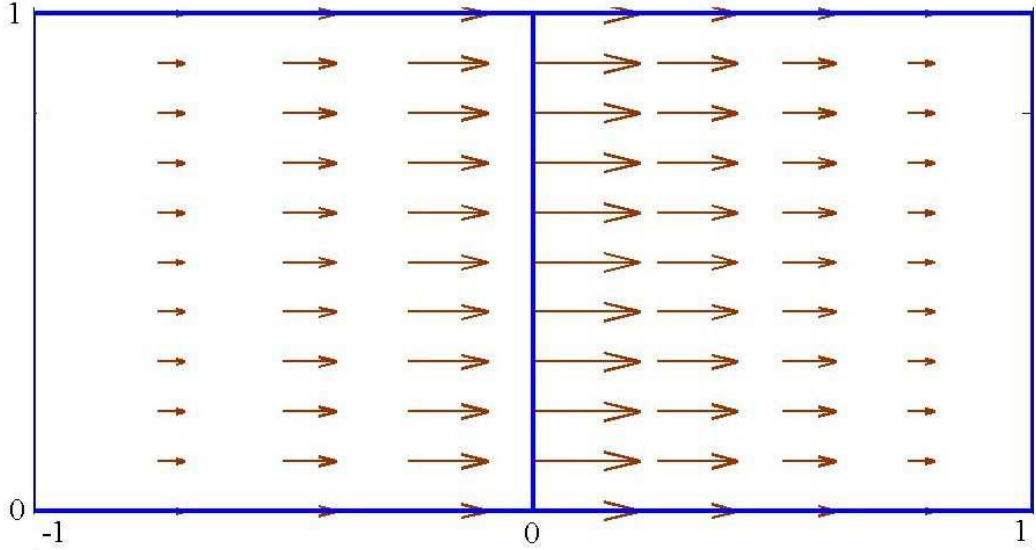}\\
  \caption{A 2D analog of the vectorial interpolation function $\bfv_\beta$
  (in this case, associated with the central vertical edge).
  Normal continuity of this function is evident from the arrow plot;
  its flux is equal to one over the central edge and zero
  over all other edges.}\label{fig:vector-face-function}
\end{figure}

The curls of $\bfw_\alpha$ are not linearly independent but rather, as can
be demonstrated, lie in the six-dimensional space spanned by the following functions:
\begin{equation}\label{eqn:hexahedral-face-interpolant}
    \bfv_{1-6} ~=~  \left\{ \frac12 (1 \pm x) \hat{x}, ~ \frac12 (1 \pm y) \hat{y},
    ~ \frac12 (1 \pm z) \hat{z} \right\}
\end{equation}
A 2D analog of a typical function $\bfv$ is shown in \figref{fig:vector-face-function}.

The $\bfv$ functions satisfy the Kronecker delta property with respect to the face fluxes:
\begin{equation}\label{eqn:hexahedral-face-interpolant-Kronecker-delta}
    [[\bfv_\beta]]_{\beta'} ~=~ \delta_{\beta \beta'}
\end{equation}
Indeed, the flux of $\bfv_1 = \hat{x}(1 + x)/2$ is obviously equal to one over the face $x = 1$
and zero over all other faces; similar conditions hold for the other five functions.
Hence one can consider vector interpolation from fluxes on the cell faces into
the volume of the cell, conceptually quite similar to edge interpolation \eqnref{eqn:hexahedral-edge-interpolation}:
\begin{equation}\label{eqn:hexahedral-face-interpolation}
    \bfq ~=~ \sum_{\beta=1}^6 \, [[\bfq]]_\beta \bfv_\beta
\end{equation}
The six-dimensional space spanned in a lattice cell
by the $\bfv$ functions will be denoted with $W_\mathrm{div}$,
reflecting the easily verifiable fact that the normal component of the field interpolated via
\eqnref{eqn:hexahedral-face-interpolation} is continuous across the common face
of two adjacent cells; hence the generalized divergence of this field exists as a regular
function, not just as a distribution (that is, there are no $\delta$-sources -- surface charges --
on the cell faces).

Importantly, as already mentioned, the curls of functions from $W_\mathrm{curl}$
lie in $W_\mathrm{div}$, or symbolically in terms of the functional spaces,
\begin{equation}\label{eqn:curlW-in-Wdiv}
    \nabla \times W_\mathrm{curl} \in W_\mathrm{div}
\end{equation}
To summarize, the following properties are critical for our construction of the coarse-grained fields:
\begin{enumerate}
  \item Twelve functions $\bfw$ (one per cell edge) interpolate the field from its circulations
  over the edges into the cell. The resulting field is tangentially continuous
  across all cell boundaries. The $\bfw$ functions span a 12-dimensional functional space
  $W_\mathrm{curl}$.
  \item Six functions $\bfv$ (one per cell face) interpolate the field from its fluxes
  through the faces into the cell. The resulting field has normal continuity
  across all cell faces. The $\bfv$ functions span a 6-dimensional functional space
  $W_\mathrm{div}$.
  \item Any vector field in $W_\mathrm{curl}$ is uniquely defined by its twelve edge circulations.
  \item Any vector field in $W_\mathrm{div}$ is uniquely defined by its six face fluxes.
  \item $\nabla \times W_\mathrm{curl} \in W_\mathrm{div}$.
 \end{enumerate}
%
\subsection{Construction of the Coarse-Grained Fields}\label{sec:EH-fields}
%
As previously noted, naive averaging of the $\bfE$ and $\bfH$ fields over the cell
adjacent to a material-air interface breaks the tangential continuity of these fields
across the boundary and is therefore unacceptable. WNBK interpolation provides
a suitable alternative. Let us define the coarse-grained fields as WNBK interpolants of the actual
edge circulations via the $\bfw$ functions, in accordance with
\eqnref{eqn:hexahedral-edge-interpolation}. Within each lattice cell,
\begin{equation}\label{eqn:E-edge-interpolation}
    \bfE ~\equiv~ \sum_{\alpha=1}^{12} \, [\bfe]_\alpha \bfw_\alpha
    ~\equiv~ \mathcal{W}_{\mathrm{curl}} ([\bfe]_{1-12})
\end{equation}
with a completely similar expression for the $\bfH$-field. The `$\equiv$' signs
indicate that this is the \emph{definition} of $\bfE$ and $\bfH$, as well as
of the WNBK curl-interpolation operator $\mathcal{W}_{\mathrm{curl}}$.

Similarly, the $\bfB$ and $\bfD$ fields are defined as interpolants in $W_\mathrm{div}$;
this interpolation, from the actual face fluxes into the cells, is effected
by the $\bfv$ functions; within each cell,
\begin{equation}\label{eqn:B-face-interpolation}
    \bfB ~=~ \sum_{\beta=1}^6 \, [[\bfb]]_\beta \bfv_\beta
    ~\equiv~ \mathcal{W}_{\mathrm{div}} ([[\bfb]]_{1-6})
\end{equation}
and a completely analogous expression for the $\bfD$ field.

We now decompose the microscopic fields $\bfe$, $\bfb$, $\bfd$
into coarse-grained parts $\bfE$, $\bfB$, $\bfD$ defined above and rapidly varying
remainders $\bfe^{\sim}$, $\bfb^{\sim}$, $\bfd^{\sim}$:
\begin{equation}\label{eqn:e-eq-E-plus-e-tilde}
    \bfe ~=~ \bfE + \bfe^{\sim}
\end{equation}
\begin{equation}\label{eqn:d-eq-D-plus-d-tilde}
    \bfd ~=~ \bfD + \bfd^{\sim}
\end{equation}
\begin{equation}\label{eqn:b-eq-B-plus-b-tilde}
    \bfb ~=~ \bfB + \bfb^{\sim}
\end{equation}
Importantly, $\bfb$ has an alternative decomposition where $\bfH$ is taken as a basis:
\begin{equation}\label{eqn:b-eq-H-plus-m-tilde}
    \bfb ~=~ \bfH + 4\pi \bfm^{\sim}
\end{equation}
With these splittings, Maxwell's equations become
\begin{equation}\label{eqn:curl-e-eq-dbdt-splitting}
    \nabla \times (\bfE + \bfe^{\sim}) ~=~ \iu \omega (\bfB + \bfb^{\sim})
\end{equation}
\begin{equation}\label{eqn:curl-B-eq-dDdt-splitting}
    \nabla \times (\bfH + 4\pi \bfm^{\sim}) ~=~ -\iu \omega (\bfD + \bfd^{\sim})
\end{equation}
It is at this point that the role of the WNBK interpolation
becomes apparent: the scales separate.
Indeed, $\bfE$ is, by construction, in $W_\mathrm{curl}$,
and therefore $\nabla \times \bfE$ is in $W_\mathrm{div}$, and so is,
by construction, $\bfB$. In that sense, the capital-letter terms
in \eqnref{eqn:curl-e-eq-dbdt-splitting} are fully compatible.
Furthermore, $\bfE$ -- again by definition -- has
the same edge circulations as the original microscopic field $\bfe$; hence,
for any face of any cell,
$$
   \int_{\mathrm{face}} (\nabla \times \bfE) \cdot \mathbf{dS}
   ~=~ \int_{\mathrm{face~edges}} \bfE \cdot \mathbf{dl}
   ~=~ \iu \omega c^{-1} \, \int_{\mathrm{face}} \bfb \cdot \mathbf{dS}
$$
where the Stokes theorem and the microscopic Maxwell equation were used.
But the $\bfB$-field has the same face fluxes as $\bfb$ by construction and, since
these face fluxes define the field in $W_\mathrm{div}$ uniquely,
we have
\begin{equation}\label{eqn:curl-E-eq-dBdt-coarse}
    \nabla \times \bfE  ~=~ \iu \omega c^{-1} \bfB
\end{equation}
Analogously,
\begin{equation}\label{eqn:curl-B-eq-dDdt-coarse}
    \nabla \times \bfH ~=~ -\iu \omega c^{-1} \bfD
\end{equation}
Thus the coarse level has separated out and, remarkably, the WNBK fields satisfy the Maxwell
equations, as well as the proper continuity conditions, \emph{exactly}.
The underlying reason for that is the compatibility of the curl- and
div-interpolations, i.e. condition \eqnref{eqn:curlW-in-Wdiv}.

For the rapidly changing components, straightforward algebra yields, from
\eqnref{eqn:curl-e-eq-dbdt-splitting} and \eqnref {eqn:curl-B-eq-dDdt-splitting}:
\begin{equation}\label{eqn:curl-e-tilde-eq-iomega-b-tilde}
    \nabla \times \bfe^{\sim} ~=~ \iu \omega \bfb^{\sim}
\end{equation}
\begin{equation}\label{eqn:curl-m-tilde-eq-iomega-d-tilde}
   4\pi \, \nabla \times \bfm^{\sim} ~=~ -\iu \omega \bfd^{\sim}
\end{equation}
with the ``constitutive relationships''
\begin{equation}\label{eqn:d-tilde-eq-eps-e-tilde-plus}
    \bfd^{\sim} ~=~ \epsilon \bfe^{\sim} + (\epsilon \bfE - \bfD)
\end{equation}
\begin{equation}\label{eqn:m-tilde-eq-b-tilde-plus}
    4\pi \, \bfm^{\sim} ~=~ \bfb^{\sim} + (\bfB - \bfH)
\end{equation}
All edge circulations of $\bfe^{\sim}$, $\bfm^{\sim}$ and all face fluxes
of $\bfb^{\sim}$, $\bfd^{\sim}$ are zero. For example,
\begin{equation}\label{eqn:e-tilde-edge-eq-0}
    [\bfe^{\sim}] ~=~ 0  ~~~ \mathrm{for~each~edge}
\end{equation}
If/when the coarse-grained fields have been found from their Maxwell equations,
one may convert the two equations for the rapid fields into
a single equation for $\bfe^{\sim}$:
\begin{equation}\label{eqn:curl-curl-e-tilde}
    \nabla \times \nabla \times \bfe^{\sim} - \omega^2 \epsilon \bfe^{\sim}
    ~=~ \omega^2 (\epsilon \bfE - \bfD) - \iu \omega \nabla \times (\bfB - \bfH)
\end{equation}
This equation can be solved for each cell, with the Dirichlet-type
zero-circulation boundary conditions \eqnref{eqn:m-tilde-eq-b-tilde-plus}.
In principle, this fast-field correction will increase the accuracy
of the overall solution. However, the remainder of the
paper will be focused on our main subject -- the coarse fields and the corresponding
effective parameters.
%
\section{Material Parameters}\label{sec:Material-parameters}
%
\subsection{Procedure for the Constitutive Matrix}\label{sec:Material-parameters-prelim}
\begin{figure}
  \centering
  \includegraphics[width=0.4\linewidth]{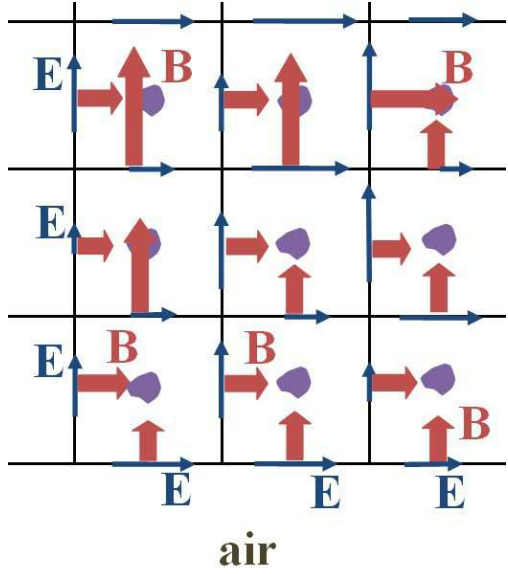}\\
  \caption{The lattice (with inclusions of arbitrary shape)
  serves as a ``scaffolding'' for the construction of coarse-grained fields.
  The curl-conforming fields such as $\bfH$ are interpolated into the cells
  from the edge circulations, while the div-conforming fields such as
  $\bfB$ are interpolated from the face fluxes.}\label{fig:field-scaffolding}
\end{figure}

%
We are now in a position to define effective material parameters,
i.e. linear relationships between $\bfD, \bfB$ and $\bfE, \bfH$.
Consider a metamaterial lattice with inclusions of arbitrary shape.
We construct a coarse-grained curl-conforming field such as $\bfH$
using the lattice cell edges as a ``scaffolding'': the field is
interpolated into the cells from the edge circulations of the respective
microscopic field (\figref{fig:field-scaffolding}). Thus, the edge circulations of the microscopic
and the coarse-grained curl-conforming fields are the same.
Similar considerations apply to div-conforming fields, but with interpolation
from the faces.

%
%
Let the electromagnetic field be approximated, in a certain region, as a linear combination of some
basis waves $\psi_\alpha$:
$$
    \Psi^{eh} ~=~ \sum_\alpha c_\alpha \psi_\alpha^{eh}; ~~~
    \Psi^{db} ~=~ \sum_\alpha c_\alpha \psi_\alpha^{db}
$$
In the most general case, $\Psi$ and all $\psi_\alpha$ are six-component vector
comprising both microscpoic fields; e.g.
$\Psi^{eh} \equiv \{ \psi^e, \psi^h \}$, etc. However, in the absence of magnetoelectric coupling
it is natural to consider each pair of fields $(\bfe, \bfd)$
and $(\bfh, \bfb)$ separately, and then the $\psi$s have three components rather than six.
The basis waves $\psi_\alpha$ can, but do not have to, be
Bloch waves. While Bloch waves are most general,
using multipole expansions with respect to a central inclusion in the lattice cell
could be beneficial in some cases.

To each basis wave, there corresponds a WNBK interpolant described above.
In the simple case of the $\bfH$ and $\bfB$ fields aligned in the same direction,
the pointiwse material parameters are found as the ratios
of $B$ and $H$ and of $D$ and $E$.

To define the material parameters in a more general setting, the field ratios
above need to be replaced with ``generalized ratios,'' as follows.
At any given point $\bfr$ in space, consider the WNBK curl-interpolants
$\bfE_\alpha(\bfr) = \mathcal{W}_{\mathrm{curl}} ([\psi^{e}_\alpha]_{1-12})$
and $\bfH_\alpha(\bfr) = \mathcal{W}_{\mathrm{curl}} ([\psi^b_\alpha]_{1-12})$
for each basis wave $\psi_\alpha$.
Similarly, consider the WNBK div-interpolants
$\bfD_\alpha(\bfr) = \mathcal{W}_{\mathrm{div}} ([[\bfd_\alpha]]_{1-6})$
and $\bfB_\alpha(\bfr) = \mathcal{W}_{\mathrm{div}} ([[\bfb_\alpha]]_{1-6})$.

Then, for each basis wave $\alpha$, we seek a linear relation
$$
     \psi_{\alpha}^{DB} (\bfr) ~=~ \zeta(\bfr) \psi_{\alpha}^{EH} (\bfr)
$$
where $\zeta$ is a $6 \times 6$ constitutive matrix characterizing,
in general, anisotropic material behavior with (if the off-diagonal blocks
are nonzero) magnetoelectric coupling. Similar to $\psi^{db}$, the six-component vectors $\psi^{DB}$
comprise both fields, but coarse-grained; same for $\psi^{EH}$.
In matrix form, the above equations are
\begin{equation}\label{eqn:Psi-DB-eq-zeta-Psi-EH}
      \Psi^{DB} (\bfr) ~=~ \zeta(\bfr) \Psi^{EH} (\bfr)
\end{equation}
where each column of the matrices $\Psi^{DB}$ and $\Psi^{EH}$ contains
the respective basis function. (Illustrative examples in subsequent
sections may help to clarify these notions and notation.)

If exactly six basis functions are chosen, one obtains the constitutive matrix
by straightforward matrix inversion; if the number of functions
is more than six, the pseudoinverse \cite{Golub96} is appropriate:
\begin{equation}\label{eqn:zeta-eq-Psi-DB-Psi-EH-plus}
      \zeta(\bfr) ~=~ \Psi^{DB}(\bfr) \, (\Psi^{EH})^+(\bfr)
\end{equation}
The approximate parameters for each cell are then found simply
by cell-averaging of $\zeta(\bfr)$. Quite a notable feature of this construction
is that the coarse-grained fields \emph{corresponding to a particular basis wave}
satisfy the Maxwell equations with this material parameter $\zeta(\bfr)$ \emph{exactly}.

One may wonder why such cell-averaging could not be done in the very beginning,
for the original ``microscopic'' Maxwell equations -- in which case one would get
the trivial value of unity by averaging the intrinsic permeability $\mu = 1$.
To see the difference, assume for simplicity that there is no magnetoelectric
coupling (this assumption does not change the essence of the argument)
and compare the curl-curl equations for the microscopic and coarse-grained fields:
$$
    \nabla \times \nabla \times \bfe \,-\,
    \left( \frac{\omega}{c} \right)^2 \epsilon(\bfr) \bfe ~=~ 0
$$
$$
    \nabla \times \zeta_\mu^{-1}(\bfr) \nabla \times \bfE \,-\,
    \left( \frac{\omega}{c} \right)^2 \zeta_\epsilon(\bfr) \bfE ~=~ 0
$$
where $\zeta_\epsilon$ and $\zeta_\mu$ are the electric and magnetic
submatrices of the $\zeta(\bfr)$ defined above. The microscopic
equation can indeed be averaged over the cell (noting that the curl
and averaging operators commute), but this would lead to the cell-averaged
field that for metamaterials is inadequate, as has already been discussed.

In the case of the coarse-grained field, the splitting of the material matrix
into its cell average plus a fluctuating component,
$\zeta = \langle \zeta \rangle + \zeta^{\sim}$, does make sense.
A perturbation analysis shows that this splitting produces
fictitious sources like $\zeta_\epsilon^{\sim}(\bfr) \bfE$
that approximately average to zero within each cell because the coarse-grained
field varies slowly and the mean of $\zeta_\epsilon^{\sim}$ is zero
by definition. The field due to these spurious equivalent sources is
therefore small. This qualitative argument is not a complete substitute for
a detailed mathematical analysis that could be undertaken in the future.
%
\subsection{The Recipe}\label{sec:Recipe}
%
Let the vectorial dimension of the problem -- i.e. the total number of vector field
components -- be $N$. Most generally in electromagnetism, $N = 6$ (three components
of $\bfE$ and three of $\bfH$), but if only one field is involved, then $N = 3$,
and if that field has only one component, then $N = 1$, etc.
Let $M$ be the total number of basis waves; since $M < N$ makes no practical sense,
we shall always assume $M \ge N$.

It is convenient to summarize the procedure described above as a ``recipe''
for finding the effective parameters:
\begin{itemize}
  \item Choose a set of basis waves $\psi_\alpha$ that provide a good approximation of
  the fields within a cell well. In general, Bloch waves are suitable candidates,
  although other choices may be possible under specific circumstances.
  The number of basis functions should be equal to or greater than the vectorial
  dimension $N$ of the problem.
  \item Find the curl- and div-WNBK interpolants of each basis wave.
  (This step requires the computation of face fluxes and edge circulations
  of the respective fields in the wave.)
  \item Assemble these WNBK interpolants into the $\Psi^{DB}$ and $\Psi^{EH}$
  matrices of \eqnref{eqn:Psi-DB-eq-zeta-Psi-EH}.
  \item Find the coordinate-dependent parameter matrix from \eqnref{eqn:zeta-eq-Psi-DB-Psi-EH-plus}.
  \item The cell-average of this matrix gives the final result: an $N \times N$
  ($6 \times 6$ in the most general case) matrix of effective parameters.
\end{itemize}

%
\subsection{Errors and Error Indicators}\label{sec:Errors}
%
Following the analysis of the previous sections, one can identify the sources
of error in the replacement of the actual metamaterial
with an effective medium. Even more importantly, ways to improve
the accuracy can also be identified, as well as the limitations of such improvement.
In the remainder, we shall ignore the usual numerical errors of evaluating
the basis waves (e.g. finite element or finite difference discretization errors
if these methods are used to compute the basis) because such errors are already
understood quite well and are only tangentially related to the subject of this paper.

Three sources of error can be distinguished in the proposed homogenization procedure.
\begin{enumerate}
  \item \emph{In-the-basis} error. If the number of basis waves is strictly greater than
  the vectorial dimension of the problem ($M > N$), then system \eqnref{eqn:Psi-DB-eq-zeta-Psi-EH}
  does not generally have an exact solution and is solved in the least-squares
  sense, as the pseudoinverse in \eqnref{eqn:zeta-eq-Psi-DB-Psi-EH-plus} indicates. From \eqnref{eqn:Psi-DB-eq-zeta-Psi-EH}
  and \eqnref{eqn:zeta-eq-Psi-DB-Psi-EH-plus}, the discrepancy between the fields is
  (with the dependence on $\bfr$ implied)
  $$
    \Psi^{DB}  - \zeta \Psi^{EH} ~=~
    \Psi^{DB} - \Psi^{DB} (\Psi^{EH})^+ \Psi^{EH} ~=~
    \Psi^{DB}(I -  (\Psi^{EH})^+ \Psi^{EH})
  $$
  where $I$ is the $6 \times 6$ identity matrix.
  Therefore the matrix norm $\| I -  (\Psi^{EH})^+ \Psi^{EH} \|$
  is a suitable  indicator of the in-the-basis error.
  If $M = N$ (for example, six basis waves in a generic problem
  of electrodynamics), then this matrix is normally zero and the in-the-basis error vanishes.
  \item \emph{Out-of-the-basis} error. Any field can be represented as a linear
  combination of the basis waves, plus a residual field. If a good basis set is chosen,
  this residual term, and hence the error associated with it, is small.
  Any expansion of the basis carries a trade-off: the residual field and
  the out-of-the-basis error will decrease, but the in-the-basis error
  may increase.
  \item \emph{Parameter averaging}. As an intermediate step, the proposed
  homogenization procedure yields a coordinate-dependent parameter matrix $\zeta(\bfr)$ that
  is in the end averaged over the lattice cell. This is discussed in the end
  of the previous section.
\end{enumerate}

The limitations of the effective-medium approximation now become apparent
as well. If a sufficient number of modes with substantially different
characteristics exist in a metamaterial (e.g. in cases of strong anisotropy),
then the in-the-basis error will be high. This is not a limitation
of the specific procedure advocated in the paper, but rather a reflection of the fact
that the behavior of fields in the material is in such a case too rich to be adequately
described by a single effective-parameter tensor.

On a positive side, several specific ways of improving the approximation
accuracy can be identified. Obviously, no accuracy gain is completely free;
better approximations may require more information than a single material
matrix can contain.
\begin{itemize}
  \item The cell problem \eqnref{eqn:curl-e-tilde-eq-iomega-b-tilde}--
\eqnref{eqn:m-tilde-eq-b-tilde-plus} for the rapidly varying fields
  can be solved, once the coarse-grained fields have been found from
  the effective-parameter model.
  \item If the relative weights of different modes in a particular model
  can be estimated \emph{a priori} at least roughly, then system \eqnref{eqn:Psi-DB-eq-zeta-Psi-EH}
  can be biased toward these modes. The downside of such biasing
  is that the material parameters are no longer a property of the material alone
  but become partly problem-dependent.
  \item The size and composition of the basis set can be optimized
  for common classes of problems, once practical experience of applying
  the methodology of this paper is accumulated.
  \item The last step of the proposed procedure -- the cell-averaging of
  matrix $\zeta$ -- can be omitted. The problem for the coarse-grained fields
  is then exact. The trade-off is that the numerical solution of such a problem
  may be computationally expensive, as the material parameter varies within the cell.
  There is of course a spectrum of practical compromises where $\zeta$
  would be approximated within a cell not to order zero (i.e. as a constant)
  but to some higher order.
  \item One may envision adaptive procedures whereby the basis waves
  are updated after the problem has been solved, and then new material
  parameters are derived from the new basis set. Recursive application
  of adaptivity may result in a substantial improvement of the solution.
  The downside, in addition to the increased cost, is the same as before:
  the material parameters become partly problem-dependent.
\end{itemize}

In connection with the last item, it is clear that the basis set
should in general reflect the symmetry and reciprocity properties
of the problem. In particular, the basis should as a rule include
\emph{pairs} of waves traveling in the opposite directions.

\subsection{Causality and Passivity}\label{sec:Causality-passivity}
%
Physical considerations indicate that the proposed procedure should be expected to
produce causal and passive effective media, at least if $M = N$. Indeed, suppose that
the opposite is true -- e.g. $\epseff'' < 0$ (with anisotropy neglected for simplicity),
violating passivity. The effective parameters, however, apply by construction
to all basis waves exactly; hence $\epseff'' < 0$ would imply power generation
in the actual physical modes in the actual passive metamaterial, which is impossible.

While such physical considerations are quite plausible, a rigorous mathematical
analysis is still desirable in future research.

%
\section{Verification}\label{sec:Verification}
%
\subsection{Empty Cell}\label{sec:Empty-cell}
%
This first test can be viewed as a ``sanity check'': will
the proposed procedure produce unit permeability and permittivity for an empty cell?
This is not a completely trivial question: it is common for the
alternative methods in the literature to yield spurious
Bloch-like factors that are then removed from the effective parameters by fiat.

For a generic plane wave passing through an empty cell in either 2D or 3D,
direct calculation that follows the proposed methodology shows
that exact parameters $\mueff = 1$ $\epseff = 1$, without any spurious factors
and of course with no magnetoelectric coupling,
are indeed obtained, regardless of the direction of the plane wave. No fudge factors or heuristic
adjustments are needed to bring these parameters to unity.

%
%

%
%

%
\subsection{Example: One-component static fields}
\label{sec:One-component-static-fields}
%
This is an obvious case that serves just as an illustrative example
and a consistency check for the proposed methodology.
Let a static field (for instance, electrostatic) have only one component (say, $z$)
that must be independent of $z$ due to the zero-divergence condition.
Then the lattice cells become effectively two-dimensional
and the div-conforming WNBK interpolant for $\bfd$ reduces just to
a constant $\bfD_0$ whose flux through the cell is equal to the flux of $\bfd$.
The curl-conforming WNBK interpolant $\mathcal{W}_{curl} (\bfE)$ would
generally be a bilinear function of $x$ and $y$,
but since the field is constant in the $xy$-plane, this interpolant also
reduces to a constant $\bfE_0 = \bfE$.

The dielectric permittivity thus is
$$
    \epseff ~\equiv~ \frac{D_0}{E_0} ~=~
    \frac{ \int_{\mathrm{cell}} \bfd \cdot \mathbf{dS} }
    {S \, E_0} ~=~ S^{-1} \, \int_{\mathrm{cell}} \epsilon \, dS
$$
exactly as should be expected.

%
\subsection{Example: Comparison with the Maxwell-Garnett formula}
\label{sec:Comparison-MG}
%
The next test is consistency with the classical Maxwell-Garnett (M-G) mixing formula for
a two-component medium, in the limit where this formula is valid.
Namely, the fill factor for the ``inclusion'' component in a host material
is assumed to be small and the field is static. (We shall not deal with radiative
corrections to the polarizability and to M-G in this paper.) The M-G expression
for the effective permittivity is, in 2D,
$$
   \epsilon_{MG,2D} ~=~ \frac{1 + f \chi} {1 - f \chi},
$$
where $\chi$ is the polarizability of inclusions in a host medium.
In particular, for cylindrical inclusions in a non-polarizable host,
$\chi = (\epsilon_{\mathrm{cyl}}-1) / (\epsilon_{\mathrm{cyl}}+1)$
and the M-G formula becomes
$$
   \epsilon_{MG,2D} ~=~ \frac{\epsilon_{\mathrm{cyl}} (1+f) + (1-f)}
   {\epsilon_{\mathrm{cyl}} (1-f) + (1+f)}
$$
The proposed methodology specializes to this case as follows.
First, one has to define a set of basis fields -- naturally,
for the static problem this is more easily done in terms of the
potential rather than the field. At zero frequency, the Bloch wavenumber is also zero;
hence the Bloch conditions for the field are periodic, but
the potential may have an offset corresponding to the line integral
of the field across the cell. Thus the first basis function $\psi_1$
is defined by
$$
   \nabla \cdot \epsilon \nabla \psi_1 \,=\, 0; ~~~
   \psi_1 \left( x = \pm \frac{a}{2} \right) = \pm  \frac{a}{2};
   ~~~~ \psi_1(y+a) - \psi_1(y) \,=\, 0
$$
The potential difference across the cell corresponds to a unit uniform
applied field in the $-x$-direction.
The second basis function is completely analogous and corresponds to a field in the $-y$-direction.
Each of the basis functions can be found by expanding it into cylindrical harmonics
$$
    \psi(r, \phi) ~=~ \sum\nolimits_{n=1}^\infty c_n g_n(r, \phi); ~~~
    g_n \equiv (r^n + p_n r^{-n}) \sin(n \phi)
$$
where the polar angle $\phi$ is measured from the symmetry line of the potential
(i.e. from the $y$ axis for $\psi_1$ and from the $x$ axis for $\psi_2$);
indexes 1 and 2 for the basis function and its respective expansion
coefficients are dropped for simplicity of notation; the potential is gauged
to zero at the center of the cell;
$p_n$ is the polarizability of the inclusion for the $n$th harmonic.
For a cylindrical particle, $p_n$ can immediately be found from the
boundary conditions on its surface:
$$
    p_{\mathrm{cyl},n} ~=~ \frac{1-\epsilon_{\mathrm{cyl}}}
    {1+ \epsilon_{\mathrm{cyl}}} \, r_{\mathrm{cyl}}^{2n}
$$
The potential at the $x$-boundaries is
$$
   \psi \left( x = \pm \frac{a}{2} \right) ~=~ \sum\nolimits_{n=1}^\infty
   c_n g_n \left( \frac{1}{2 \sin \phi}, \, \phi \right)
$$
The coefficients $c_n$ can be found numerically by truncating the infinite series
at $n = n_{\max}$ and applying the Galerkin method. This leads to a system of equations
$$
    G \underline{c} ~=~ \underline{f}
$$
Here $\underline{c}$ is the Euclidean vector of the expansion coefficients,
$$
   G(n, m) = \int_{x = a/2} g_n g_m dl; ~~~
   f_n = \frac12 \, \int_{x = a/2} g_n dl
$$
The system of equations may be solved numerically.

Once the expansion coefficients and hence the basis functions
have been found, the next steps are to compute the circulations and fluxes, and
after that the WNBK interpolants, of the basis functions.
For $\psi_1$, the circulation of the respective field $\bfe_1 = -\nabla \psi_1$
over each ``horizontal'' edge $y = \pm \frac12$ is, by construction, equal to $a$
and is zero over the two ``vertical'' edges. The fluxes of $\bfd_1 = \epsilon \bfe_1$
are zero over the horizontal edges; for the vertical ones,
$$
    [[\bfd_1]] ~=~ \int_{y = \pm a/2} d_{1x}  dy
    ~=~ -\int_{y = \pm a/2} \epsilon \, \partial_x \psi_{1}  dy
$$
where $\epsilon$ is that of the host if cell boundaries do not cut
through the inclusions; the $x$-derivative can be computed from
the $r$- and $\phi$-derivatives of the cylindrical harmonics.

Since the fluxes and circulations of the fields over the pairs of opposite
edges are in this case equal, the WNBK interpolant is simply constant
and the effective parameters are obtained immediately, without
the intermediate stage of their pointwise values. Formally,
the system of equations for $\epseff$ is
$$
    \begin{pmatrix}
    1 & 0\\
    0 & 1
    \end{pmatrix} \,
    \begin{pmatrix}
    \epsilon_{xx} & \epsilon_{xy} \\
    \epsilon_{yx}  & \epsilon_{yy}
    \end{pmatrix} ~=~
    \begin{pmatrix}
    [[d_1]] & 0 \\
    0  & [[d_2]]
    \end{pmatrix}
$$
the identity matrix is written out explicitly to clarify its origin:
its first column represents the WNBK curl-conforming interpolant of $\bfe_1$; the second column --
that of $\bfe_2$. The rightmost matrix contains the div-conforming WNBK interpolants
of $\bfd_1$ and $\bfd_2$. As can be expected, the system of equations is in this
case trivial. Since the fluxes $[[\bfd_1]]$ and $[[\bfd_2]]$ are equal,
the permittivity is, as expected, a scalar quantity equal to these fluxes.

\begin{figure}
  \centering
  \includegraphics[width=0.55\linewidth]{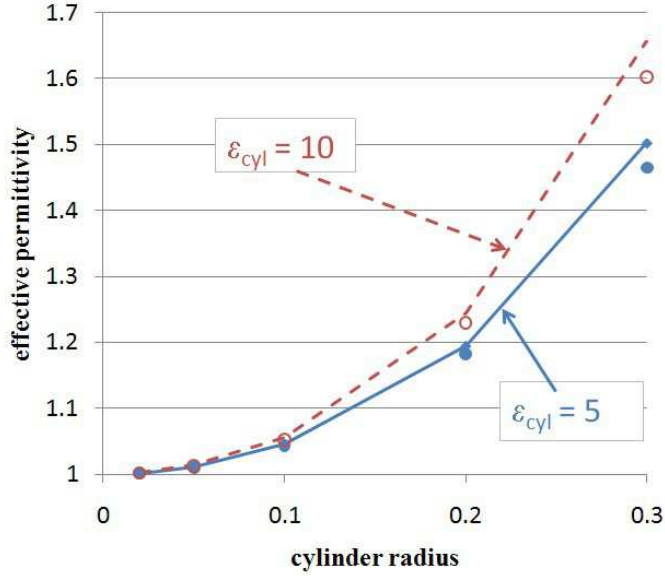}\\
  \caption{Effective $\epsilon$ for a 2D-periodic array of cylinders.
  Lines: the proposed procedure (with 9 cylindrical harmonics).
  Markers: the Maxwell-Garnett formula.}\label{fig:eps-eff-PhC-vs-r0}
\end{figure}

For numerical illustration, let us consider a cylindrical inclusion
with a varying radius.
The plots of $\epseff$ vs radius of the cylinder
(\figref{fig:eps-eff-PhC-vs-r0}) for two different values of the permittivity
($\epsilon_\mathrm{cyl} = 5$ and $\epsilon_\mathrm{cyl} = 10$)
show an excellent agreement between the new method and M-G,
in the range where such an agreement is to be expected -- that is, for small
radii of the inclusion.

%
\subsection{Example: Bloch Bands and Wave Refraction}
\label{sec:Wave-refraction}
%
Definitive tests of the effective parameters come from Bloch band diagrams
and, even more importantly, from wave reflection and refraction at material interfaces. Here we consider
wave propagation through a photonic crystal (PhC) slab -- an array of
cylindrical rods with no defects. For consistency with previous work,
the geometric and physical parameters of the crystal are chosen to be the same
as in the tests reported previously \cite{Tsukerman-PBG08} and are taken
from \cite{Gajic05}. Namely, the radius of the rod is $r_{\mathrm{cyl}} = 0.33 a$
and its dielectric permittivity is $\epsilon_{cyl} = 9.61$.
The $p$-mode ($H$-mode, with the one-component magnetic field along the rods)
is considered because this is a more interesting case for homogenization.

The numerical simulation of the Bloch bands and of wave propagation
through the PhC slab was performed with FLAME \cite{Tsukerman-PBG08,Tsukerman-book07},
a generalized finite difference calculus that is, arguably, the most accurate
method for this type of problem because it incorporates local analytical solutions
into the difference scheme.

\begin{figure}
  \centering
  \includegraphics[width=0.55\linewidth]{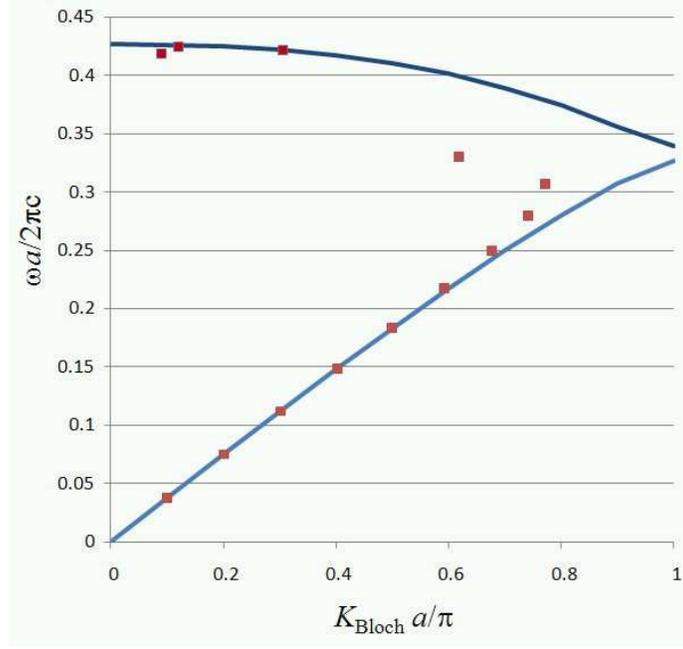}\\
  \caption{The $\Gamma X$ Bloch bands obtained with the effective parameters
   (markers) vs. accurate numerical simulation (solid line).
   $\epsilon_{cyl} = 9.61$, $r_{cyl} = 0.33 a$.}\label{fig:Gajic-PhC-GX-Bloch-bands}
\end{figure}

\figref{fig:Gajic-PhC-GX-Bloch-bands} shows that the Bloch bands obtained with the effective parameters
are in an excellent agreement with the accurate numerical simulation, except for
a few data points at the band edge, where effective parameters cannot be expected
to remain valid.
 
\begin{figure}
  \centering
  \includegraphics[width=0.55\linewidth]{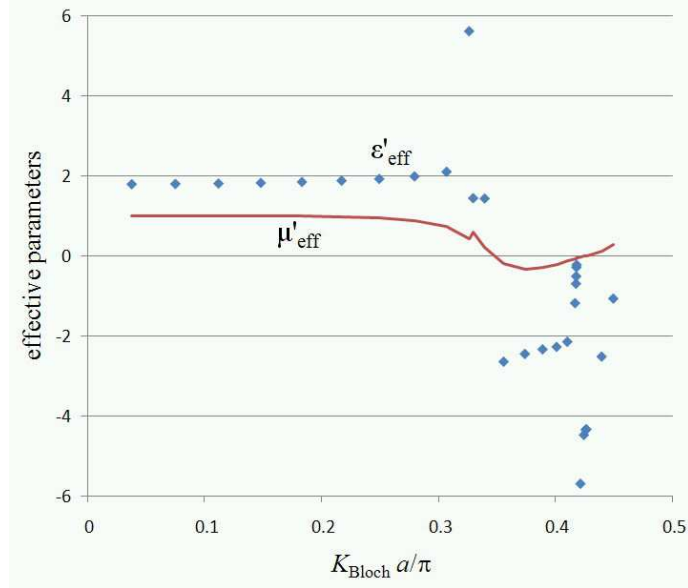}\\
  \caption{Effective parameters (markers -- $\epseff'$, solid line -- $\mueff'$)
   vs the Bloch wavenumber in the $\Gamma X$ direction. $\epsilon_{cyl} = 9.61$, $r_{cyl} = 0.33 a$.}\label{fig:Gajic-PhC-eff-params-vs-K}
\end{figure}

\figref{fig:Gajic-PhC-eff-params-vs-K} displays the real parts of $\epseff$ and $\mueff$ as a function
of the Bloch wavenumber, in the $\Gamma X$ direction. Among other features,
a region of double-negative parameters ($\epseff < 0$, $\mueff < 0$)
can be clearly identified for the normalized Bloch wavenumber
approximately between 0.33 and 0.42. This agrees very well with the
band diagrams above and with the results reported previously \cite{Gajic05,Tsukerman-PBG08}.

\begin{figure}
  \centering
  \includegraphics[width=0.55\linewidth]{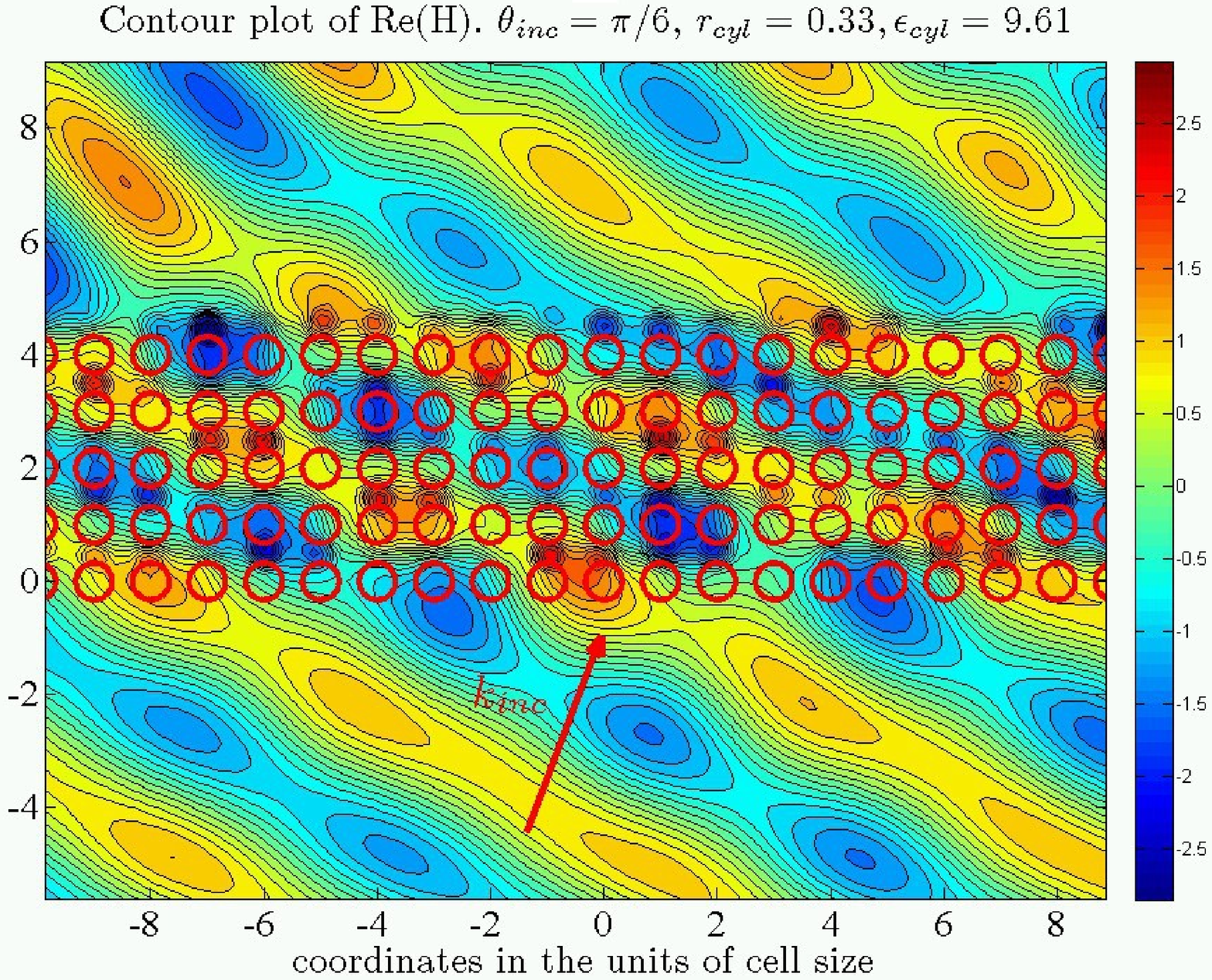}\\
  \caption{Contour plot of Re($H$). Angle of incidence  $\pi / 6$, $\epsilon_{cyl} = 9.61$, $r_{cyl} = 0.33 a$.}\label{fig:Gajic-PhC-Re-H-contourplot}
\end{figure}

Numerical simulation (using FLAME) of EM waves in the actual PhC is compared with
the analytical solution for a homogeneous slab of the same thickness as the PhC
and with effective material parameters $\epseff$ and $\mueff$ calculated as
the methodology of this paper prescribes. In the numerical simulations, obviously, a finite array
of cylindrical rods has to be used -- in this case, $24 \times 5$ to limit the computational cost.
The results reported below are at the normalized frequency of $\tilde{\omega}
\equiv \omega a / (2\pi c) = (\lambda/a)^{-1} = 0.24959$, coinciding with
one of the data points in previous simulations \cite{Tsukerman-PBG08}.
A fragment of the contour plot of Re($H$) in \figref{fig:Gajic-PhC-Re-H-contourplot} helps
to visualize the wave for the angle of incidence $\pi /3$.

\begin{figure}
  \centering
  \includegraphics[width=\figurewidth]{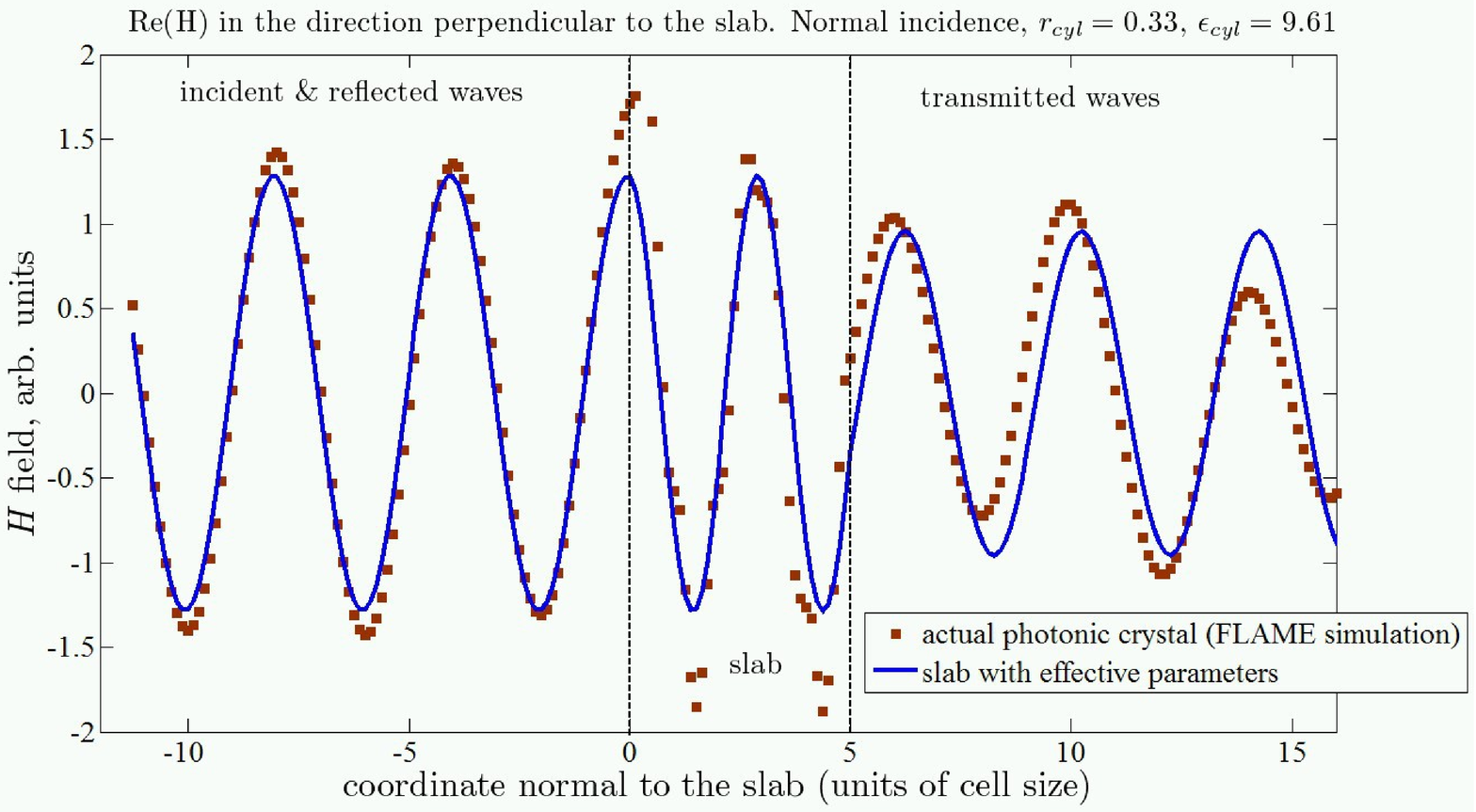}\\
  \caption{Re($H$) along the coordinate perpendicular to the slab. Normal incidence.
  Other parameters same as above. As expected, the ``microscopic'' field (markers) exhibits
  some scatter within the slab, and nevertheless agrees fairly well with the effective
  medium field (solid line). The reasons for soem discrepancy are noted in the text.}\label{fig:Gajic_PhC-Re-H-vs-coordinate-normal-incidence}
\end{figure}

\begin{figure}
  \centering
  \includegraphics[width=\figurewidth]{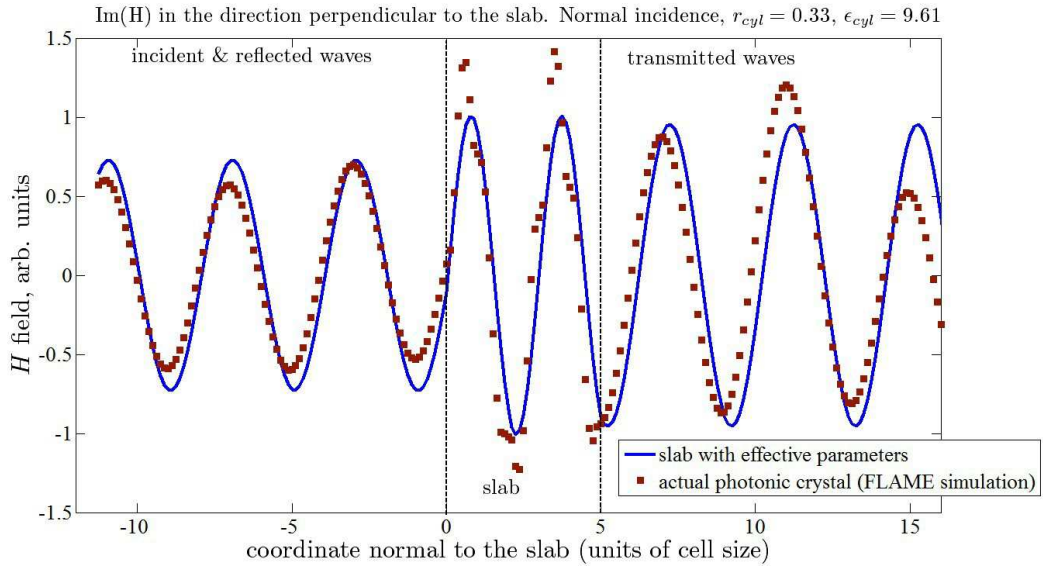}\\
  \caption{Im($H$) along the coordinate perpendicular to the slab. Normal incidence. $\epsilon_{cyl} = 9.61$, $r_{cyl}/a = 0.33$.}\label{fig:Gajic_PhC-Im-H-vs-coordinate-normal-incidence}
\end{figure}

\begin{figure}
  \centering
  \includegraphics[width=\figurewidth]{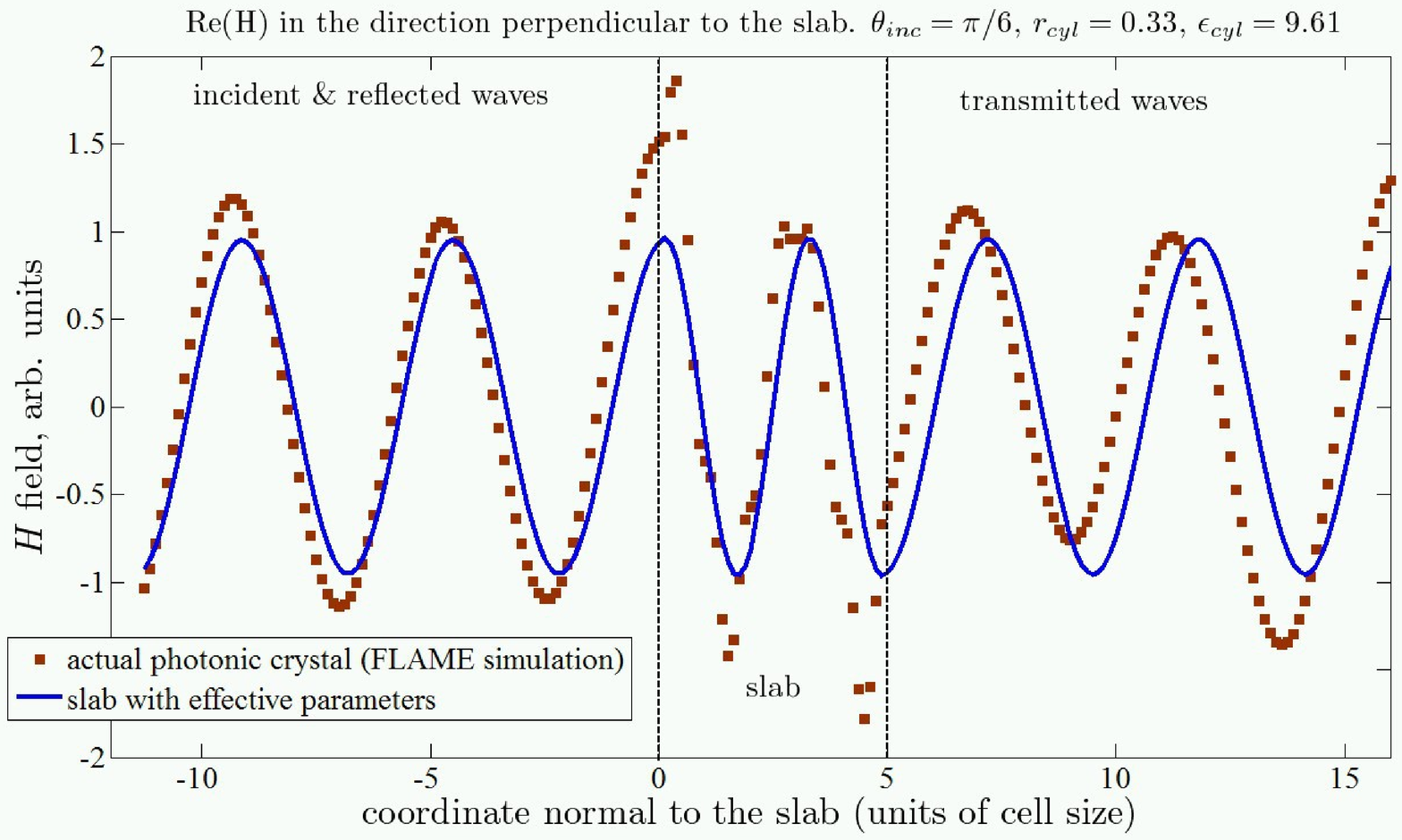}\\
  \caption{Re($H$) along the coordinate perpendicular to the slab. The angle of incidence
  $\pi / 6$. Other parameters same as above.}\label{fig:Gajic-PhC-Re-H-vs-coord-inc-angle-pi6}
\end{figure}

\begin{figure}
  \centering
  \includegraphics[width=\figurewidth]{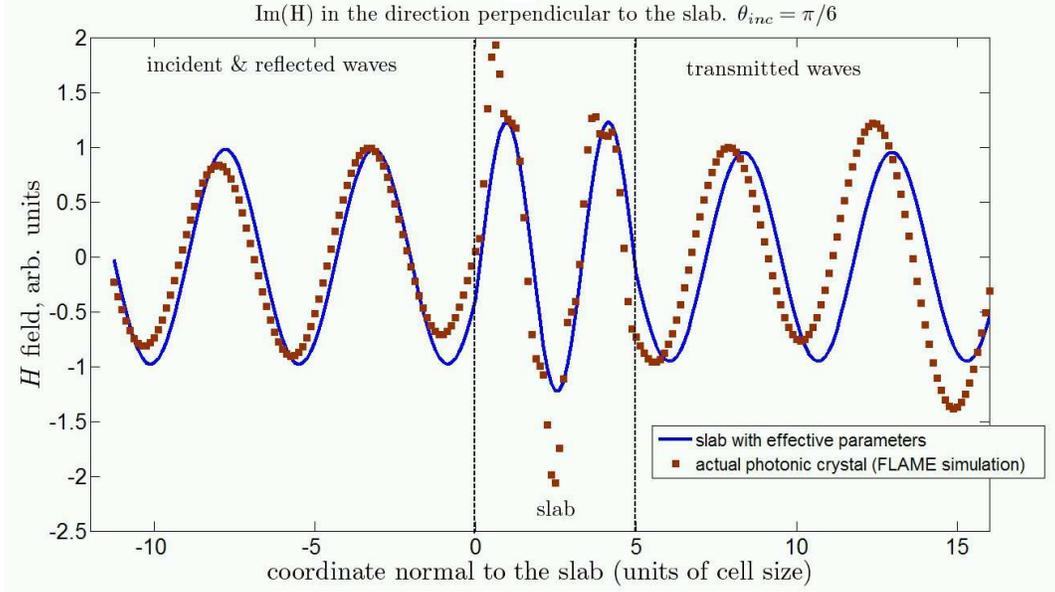}\\
  \caption{Im($H$) along the coordinate perpendicular to the slab. The angle of incidence
  $\pi / 6$. Other parameters same as above.}\label{fig:Gajic-PhC-Im-H-vs-coord-inc-angle-pi6}
\end{figure}

The real and imaginary parts of the numerical and analytical magnetic fields
along the line perpendicular to the slab are plotted in Figs.~
\ref{fig:Gajic_PhC-Re-H-vs-coordinate-normal-incidence} and
\ref{fig:Gajic_PhC-Im-H-vs-coordinate-normal-incidence}, respectively, for normal
incidence and in Figs.~\ref{fig:Gajic-PhC-Re-H-vs-coord-inc-angle-pi6}
and \ref{fig:Gajic-PhC-Im-H-vs-coord-inc-angle-pi6} for the angle of incidence $\pi / 6$.
The analytical and numerical results are seen to be in a good agreement;
some discrepancies can be explained by (i) numerical artifacts due to the finite length
of the PhC in the tangential direction; (ii) the finite thickness of the slab
that makes it different from the bulk; (iii) the approximate nature of the effective
parameters, especially in this case where the cell size is $\sim$1/4 of the wavelength
and the filling factor is high.

%
\subsection{Conclusion}
\label{sec:Conclusion}
%
A new methodology is put forward for the evaluation of the effective parameters
of electromagnetic and optical metamaterials. The main underlying principle
is that the coarse-grained $\bfE$ and $\bfH$ fields have to be curl-conforming
(that is, they have to possess a valid curl as a regular function,
which in particular implies tangential continuity across material interfaces),
while the $\bfB$ and $\bfD$ fields have to be div-conforming, with their
normal components continuous across interface boundaries. While some flexibility
in the choice of these coarse-grained fields exists, an excellent framework
for their construction is provided by Whitney forms and the WNBK complex (\sectref{sec:WNBK-complex}).
This construction ensures not only the proper continuity conditions
for the respective fields, but also the compatibility of the respective
interpolants, so that e.g. the curl of $\bfE$ lies in the same approximation space
as $\bfB$. As a result, remarkably, Maxwell's equations for the coarse-grained fields
are satisfied \emph{exactly}.

Further, the electromagnetic field is approximated with a linear combination
of basis functions, the most general choice of which is Bloch waves,
although in special cases other options may be available. Effective parameters are
then devised to provide the most accurate linear relation between the WNBK
interpolants of the basis functions. In the limiting case of a vanishingly small
cell size, this procedure yields the exact result; for a finite cell size --
as must be the case for all metamaterials of interest \cite{Merlin09,Tsukerman08} --
the effective parameters are an approximation.
Ways to improve the accuracy are outlined in \sectref{sec:Errors}.

Proponents of the differential-geometric treatment of electromagnetic theory
have long argued that the $\bfH$, $\bfE$ and $\bfB$, $\bfD$ fields are actually
different physical entities, the first pair best characterized via circulations
(mathematically, as 1-forms) and the second one via fluxes (2-forms)
\cite{Tonti72,Kotiuga85,Bossavit98}. The approach advocated and
verified in this paper buttresses this viewpoint.

There are several important issues to be addressed in future work.
First, the theory of this paper needs to be applied to
3D structures of practical interest. Second, physical considerations indicate
that the theory should lead to causal and passive
effective media (see \sectref{sec:Causality-passivity}),
but a rigorous mathematical analysis or, in the absence of that,
accumulated numerical evidence are highly desirable. The significance
of passivity and causality is evident and has been emphasized
by Simovski \& Tretyakov \cite{Simovski09,Simovski-Tretyakov10}.

Although the object of interest in this paper is artificial metamaterials,
it is hoped that the new theory will also help to understand more deeply the nature of
the fields \emph{in natural materials}\, because rigorous definitions of such fields,
especially of the $\bfH$ field, are nontrivial. The ideas and methodology of the paper
are general and should find applications beyond electromagnetism,
for example in acoustics and elasticity.

\section*{Acknowledgment}
%
Discussions with Vadim Markel and John Schotland
inspired this work and are greatly appreciated. Communication with
Alain Bossavit, C.~T.~Chan,  Zhaoqing	Zhang, Ying Wu, Yun Lai is also gratefully acknowledged.
I thank C.~T.~Chan (HKUST, Hong Kong), W.~C.~Chew \& Lijun~Jiang (HKU, Hong Kong),
Oszk\'{a}r B\'{i}r\'{o}, Christian Magele, Kurt Preis (TU Graz, Austria) and
Sergey Bozhevolnyi (Syddansk Universitet, Denmark) for their hospitality
during the period when this work was either contemplated or performed.

\bibliographystyle{plain}


\end{document}